# Thermodynamics of the space of trajectories governed by a combination of two additive boundary functionals

V. V. Ryazanov

Institute for Nuclear Research, pr. Nauki, 47 Kiev, Ukraine, e-mail: vryazan19@gmail.com

This paper extends the formalism of stochastic path-space thermodynamics by systematically expanding the space of thermodynamic variables with a spectrum of boundary and relative functionals of random processes. Generalizing the approaches introduced in preprints arXiv:2607.24078 and arXiv:2609.08477, we investigate the fluctuation statistics of a one-dimensional Ornstein–Uhlenbeck process with an asymmetric linear drift and a step penalty potential at the origin, which models the waiting phase of a molecular Brownian motor. Utilizing the Feynman–Kac formalism and the parabolic cylinder theory, a numerical scheme based on matching logarithmic derivatives via the secant method is constructed. A high-precision eigenvalues of the effective Hamiltonian is obtained, which define the cumulant generating function. We formulate a modified path-space fluctuation theorem of the Gallavotti–Cohen type for the conjugate drift current at a fixed integral occupation time of the system in the dissipative half-plane. The physical and thermodynamic significance of the new variables is elucidated. A combination of "residence time" and "integral current" functionals is applied to describe an ion channel sensitive to mechanical or electrical stimuli.

## 1. Introduction

In recent decades, stochastic thermodynamics and large deviation theory [1–5] have established themselves as a fundamental mathematical framework for describing open systems far from thermodynamic equilibrium. They have brought about a qualitative breakthrough in our understanding of the physics of non-equilibrium systems. The classical approach to formulating fluctuation theorems and equations of state for non-equilibrium processes—which traditionally relies on instantaneous phase-space coordinates or integral quantities such as total work, heat transfer, and entropy production over finite or infinite time intervals (as well as state variables like temperature, internal energy, or chemical potential)—proves insufficient for describing microscopic and open systems. For mesoscopic objects—including biological macromolecules, nanoscale motors, quantum nanodevices, and artificial open systems, where fluctuations are comparable in magnitude to average values—it is not only integral averages that are crucial, but also the geometric invariants of the shapes of random trajectories. This has necessitated a shift from the thermodynamics of states to the thermodynamics of trajectories (path-space thermodynamics) [6–7], wherein the object of study becomes the shape of the fluctuating trajectory over a finite or asymptotic time interval.

Works [8–12] proposed a radical extension of the standard thermodynamic state space by incorporating extremal boundary functionals—specifically, the first-passage time to a given level. This approach enabled the coupling of stochastic stopping times with generalized thermodynamic forces, revealing new kinetic aspects of fluctuation phenomena. This formalism was further developed in preprint [13], where the space of controlling trajectory parameters was augmented to include the process maximum over a given interval and the system's occupation time above a fixed energy barrier. This extension linked the geometry of the fluctuation corridor to integral dissipation and laid the groundwork for deriving new classes of fluctuation theorems that account for the "trajectory memory" of open systems; this allows for the coupling of extremal geometric fluctuation characteristics with generalized thermodynamic forces (such as chemical potentials, intensities, or penalties), thereby establishing a formalism of thermodynamics in trajectory space (Path-space thermodynamics).

Despite the significant conceptual importance of these results, the question of the mutual influence and cross-correlations among several disparate trajectory invariants has remained unexplored in the literature. Most real-world mesoscopic and biological systems operate in multiphysic fields where the kinetics of switching between discrete conformational states (characterized by residence times) is coupled with continuous mass or charge transport (characterized by integral currents). A prime example is found in mechano- and voltage-gated ion channels in biological membranes, where the open-state duration of the gating mechanism is stochastically correlated with the total ionic current flowing through the fluctuating pore geometry.

In this work, we investigate for the first time the thermodynamics of the space of trajectories governed by a combination of two additive boundary functionals: the process's residence time in the positive half-plane, $\tau_+$, and the integral current (the position of the trajectory's center of mass), $I_T$. Within the framework of a stochastic Langevin equation modeling an asymmetric Ornstein–Uhlenbeck process with a fundamental energy gap $U_0$, we derive an exact thermodynamic potential (the dynamic free energy of large deviations) using the Feynman–Kac formalism. An analytical solution to the corresponding stationary Schrödinger equation with a piecewise linear potential is obtained by matching parabolic cylinder functions (Weber functions). We justify extending the thermodynamic space to include this pair of additive functionals (residence time and current) in order to describe the history of non-equilibrium processes.

The main result of the work is the derivation of an exact expression for the mutual cross-sensitivity coefficient $\chi_{12}$, which serves as a trajectory-based analogue of the Onsager reciprocity relations for systems with broken spatial symmetry. We demonstrate the existence of a nonlinear "sensitivity window" and investigate the stabilizing role of thermal noise, which activates a coherent system response to the combined action of chemical and electrical macroscopic forces.

The scientific novelty of this work lies in taking the next step toward expanding the trajectory phase space. While previous studies focused on integral characteristics (residence time) or simple extrema (maxima), this article introduces and investigates—for the first time—a comprehensive set of differential and relative boundary functionals that extract more nuanced information about the structure of the non-equilibrium process. These new variables may include:

a). Maximal Deficit (Maximum Drawdown), characterizing the process's maximum drop relative to its historical peak; physically, this functional defines an upper bound for the fluctuation-induced violation of the Second Law of Thermodynamics over finite time intervals (system "fatigue"). b). Time spent in the bankruptcy/negative-value half-plane ($x < 0$), which strictly determines the net fraction of time the system spends in an adverse or dissipative state, thereby allowing for a rigorous distinction between thermodynamic phases along a single individual trajectory. c). Time of return following bankruptcy, characterizing the system's recovery time after a deep critical fluctuation, as well as other boundary-related functionals of stochastic processes.

Each of the specified functionals reveals a unique physic-thermodynamic significance, translating continuous stochastic thermodynamics into a discrete-event framework. Employing the Feynman–Kac formalism and parabolic cylinder theory, the authors have constructed a generalized cumulant generating function (energy spectrum) for a diffusion process characterized by linear drift and a step-like penalty, and have formulated a modified fluctuation theorem of the Gallavotti–Cohen type. These results open up new possibilities for the precise calculation of catastrophic drawdown probabilities and the optimization of control protocols under intense fluctuation regimes.

The article is organized as follows. Section 2 describes how the use of boundary functionals—with their physical significance specified—extends the space of thermodynamic variables. Section 3 outlines the mathematical framework (the Feynman-Kac method for the Ornstein-Uhlenbeck process). Section 4 presents an example of applying the proposed approach to a microscopic biological system. Section 5 discusses the results, and Section 6 provides the conclusion. Appendix A addresses mathematical issues, while Appendix B presents free-energy profiles.

## 2. Extension of the space of thermodynamic variables via boundary (extremal) functionals of random processes. Physical-thermodynamic significance of the boundary functionals.

Expanding the space of thermodynamic variables to include boundary (extremal) functionals of stochastic processes [8–13] represents a development of the formalism for the thermodynamic description of nonequilibrium systems (in the spirit of the coarse-grained description of fluctuations, large deviation theory, and stochastic thermodynamics [1–7]).

Introducing the first-passage time [8–12], the maximum, and the occupation time [13] makes it possible to relate these stochastic invariants to the corresponding generalized forces (intensities, chemical potentials, or penalties). Incorporating other classical boundary functionals reveals their physical-thermodynamic significance. Let us consider examples involving several such functionals.

1. Maximum Deficit (Maximum Drawdown). This metric $D_T = \max_{0\le t\le T}(\sup X_s - X_t)$ measures the maximum decline of the process relative to its historical maximum over a given interval.

Physical meaning in thermodynamics: it describes the maximum fluctuation-induced loss (dissipation) or "fatigue" of a system along a trajectory. If $X_t$ is associated with free energy or entropy, then the maximal deficit sets an upper bound on the fluctuation-induced violation of the second law of thermodynamics over finite time intervals.

A thermodynamic potential governing system resilience is introduced. The variable conjugate to the deficit acts as an "insurance potential"—or a measure of irreversibility—controlling the probability of catastrophic drops (critical fluctuations) in open systems.

2. Duration of the process's stay in the half-plane $X_t \le 0$ of bankruptcy/negative values. This functional (related to distributions of the Lévy arcsine law type) captures the net proportion of time the system spends in an adverse or "forbidden" state.

In thermodynamics, this functional characterizes metastability and phase separation. In stochastic thermodynamics, residing in the region $X_t \le 0$ can signify the machine operating in an energy-consuming mode rather than a generating one (or vice versa).

The introduction of this functional allows for a strict distinction between thermodynamic phases along a single trajectory. The conjugate potential (analogous to the chemical potential of a phase) determines the "penalty" for residing in a metastable or dissipative state. This provides a tool for describing dynamic phase transitions, where the governing parameter is not an instantaneous value but the integral lifetime of the structure.

3. The moment of return after bankruptcy $\tau_{ret}$. This is the time of the first return to the positive half-plane after the process has crossed zero in the downward direction. In thermodynamics, it characterizes the relaxation (or recovery) time of a system following a deep fluctuation. Incorporating this moment as a variable enables the introduction of the concepts of "thermodynamic memory" and reversibility. In standard Markovian thermodynamic systems, the return time is implicitly embedded in the evolution operator. Explicit coupling $\tau_{ret}$ with the thermodynamic force makes it possible to describe processes involving memory effects (non-Markovian properties of the effective subsystem) and to derive fluctuation theorems for degradation-recovery cycles.

4. Other functionals (e.g., the number of level crossings / number of upcrossings). Physical significance: counting the number of cycles or oscillations of the system around a steady state. These functional transforms continuous thermodynamics into a discrete-event framework. The variable conjugate to the number of crossings is the effective frequency (or action "quantum"). This is critically important for describing stochastic motors (molecular machines) and biochemical oscillators, where the key factor is not merely the trajectory, but the number of successfully completed work cycles per unit of time.

Incorporating these functionals transforms the standard phase space ("coordinates–momenta–volumes–entropies") into a space of trajectory-shape characteristics (path-space thermodynamics). This enables: a) the construction of generalized equations of state based on the process history rather than instantaneous macro-parameters ("historical thermodynamics"); b) the derivation of new fluctuation relations (Fluctuation Theorems) for extreme values—crucial for microscopic systems (biopolymers, quantum dots, nanomachines) where average values provide little information due to massive fluctuations; and c) the optimization of control protocols in stochastic thermodynamics (e.g., minimizing the maximum deficit at a fixed average power).

Generating functions or large deviation functions for these functionals are derived for various classes of processes (diffusion processes, Lévy processes, fractional Brownian motion).

Each boundary functional possesses physical significance and opens up new possibilities for describing thermodynamic systems; each new functional defines a new "dimension" in the space of trajectories, transforming a random trajectory into a point in a new thermodynamic space.

A strict hierarchy of functional "importance" can be constructed by selecting as a criterion the degree of information coarsening (granularity) that the functional extracts from the trajectory, as well as its connection to fundamental concepts (entropy, work, time symmetry).

Presented below is the thermodynamic hierarchy of boundary functionals—ranging from the basic (integral) to the most subtle (extremal).

Level 1: Integral invariants of the trajectory (Foundations). Examples: Occupation time above/below a level, Feynman-Kac type integral functionals ($\int_0^T V(X_t)dt$). They rank first because they possess the property of time-additivity. If a trajectory is partitioned into segments, these functionals simply sum up. They hold paramount thermodynamic significance. They are directly linked to time averages, ergodicity, and classical fluctuation theorems (such as those of Gallavotti-Cohen and Jarzynski). They measure the fraction of time spent in a given phase, a value that translates directly into classical thermodynamic weights.

Level 2: Topological markers and stopping times (boundaries and events). Examples: first passage time, time of return after bankruptcy. Ranked second because these are Markovian stopping times. They are not additive but possess the strong Markov property (the process "forgets" the past at that moment). Thermodynamic significance: Critical for open and metastable systems. First passage time determines the lifetime of a metastable state or the time until thermal breakdown or a chemical reaction (Kramers kinetics). These functionals link thermodynamics with kinetics and transport processes.

Level 3: global extrema (geometry of the fluctuation corridor). Examples: the process maximum/minimum over an interval ($\max X_t$). It ranks third because these are essentially non-local and non-additive functionals. Determining the maximum requires examining the entire trajectory; a time increment does not guarantee a change in the maximum. It holds high thermodynamic significance for reliability theory and the study of rare events. The maximum defines stability limits. In thermodynamics, it sets the "ceiling" for fluctuations (for instance, the maximum instantaneous work a fluctuation can perform against a field).

Level 4: Differential and relative extrema (fine structure). Examples: maximal deficit, number of levels upcrossings. Why Level 4: they depend not merely on the process values, but on the history of the relationship between the current value and a past extremum (as in the case of deficit). Their thermodynamic significance is specialized. They are important for systems exhibiting memory, hysteresis, or aging, as well as for feedback-controlled systems (information thermodynamics / Maxwell's demons). They describe the "microstructure" of dissipation.

Summary hierarchy table. The degree of importance of the functionals is determined by the level of information aggregation and their mathematical properties:

| Level/Rank | Class of functionals | Mathematical property | Example of application in physics | What it measures in physics |
|---|---|---|---|---|
| 1. Higher | Integral (Residence time) | Time additivity, Markov property | Definition of macrostates, fluctuation theorems | Phase statistical weight, total dissipation |
| 2. Basic | Local times (first hitting, return) | Stopping times, strong Markov property | Kramers reaction kinetics, phase transitions, thermal breakdown | State lifetime, relaxation rate |
| 3. Intermediate | Global extrema (Maximum / Minimum) | Significant non-locality, non-additivity | Calculation of structure failure probability, extreme fluctuations | Limits of fluctuation stability |
| 4. Subtle | Relative extrema (Maximal Deficit) | Dependence on history and historical trend | Thermodynamics of systems with adaptation, aging, and feedback | Memory effects, fatigue, hysteresis |

Conclusion. In this context, "importance" is equivalent to universality. Residence time and first-passage time are important for any thermodynamic system. However, the maximum deficit becomes a critically important tool in cases where the system adapts or undergoes wear, or where the trajectory-based cost of a "drop" from a historical peak is significant.

To mathematically characterize this hierarchy, researchers typically examine the joint distribution of these functionals—for instance, the relationship between the maximum value and the time at which it is attained. Proposing the use of a combination of two additive (integral) functionals is an excellent and highly robust strategic move. Within the hierarchy of thermodynamic variables, additivity is prized above all else, as it enables the direct application of the powerful framework of Large Deviation Theory and preserves a structure analogous to classical thermodynamics (where entropy and energy are likewise additive).

We will consider a combination of occupation time and the integral functional of the current (or entropy production). Option 1: "Energy + Efficiency" (classical phase choice). Two functionals are considered:

$$1.\ \tau_+ = \int_0^T \Theta(X_t)dt \tag{1}$$

is the time the process spends in the positive half-plane (above level 0), where Θ(x) is Heaviside function.

$$2.\ I_T = \int_0^T f(X_t)dt \ \text{ (or } I_T = \int_0^T f(X_t)dX_t,\ I_T = \int_0^T X_t dt \text{)} \tag{2}$$

is integral current, work performed, or accumulated energy over the entire time T.

Using such a combination makes it possible to describe the system in terms of "where we were" and "how much we earned or lost in the process."

This offers the following advantage for the theory: it is possible to construct a generalized free-energy-like potential $G(\lambda_1,\lambda_2)$ conjugate to both functionals. The conjugate force $\lambda_1$ acts as a "phase chemical potential," while $\lambda_2$ acts as an inverse temperature. This is an ideal model for describing dynamic phase separation in microscopic motors (for example, when a motor switches between efficient and inefficient operating modes).

One may also consider combinations such as: a) Variant 2, two additive functionals for two distinct zones: $\tau_A = \int_0^T \Theta(X_t - a)dt$ — time spent above the high level *a* (zone of extreme activation), $\tau_B = \int_0^T \Theta(-X_t - b)dt$ — time spent below the low level *b* (zone of deep deficit); b) Variant 3, "Local Time + Global Trend." Instead of time spent in a half-plane, the following are used: $L_T^0$ — local time at zero (the integral $\int_0^T \delta(X_t)dt$; although formally a singular functional, it possesses the property of time-additivity and measures "viscosity" or the frequency with which the process returns to the initial/equilibrium state, $I_T = \int_0^T X_t dt$ — the average integral position (center of mass of the trajectory).

### 3. Mathematical Framework: The Feynman-Kac Method for the Ornstein-Uhlenbeck Process

Let us consider the classical Ornstein-Uhlenbeck (OU) process defined by the stochastic Langevin equation: : $dX_t = -\gamma X_t\, dt + \sqrt{(2D)}\, dW_t$, where γ is the restoring force coefficient, D is the diffusion coefficient related to the heat bath temperature $T_{th}$ via the Einstein relation, and $W_t$ is a standard Wiener process. To construct the thermodynamic potential, we select a two-parameter combination of additive functionals (Option 1): the residence time in the positive half-plane (1) and the integrated current (mean position) (2).

Why is the combination of two additive functionals effective?

1. Mathematical rigor (Kac's theorem): For two additive functionals $A_1 = \int V_1(X_t)dt$ and $A_2 = \int V_2(X_t)dt$, the joint Laplace transform is calculated via a generalized Schrödinger equation (or a modified Fokker-Planck equation):

$$\mathcal{L}\psi = \left(\hat{L}_{FP} - \lambda_1 V_1(x) - \lambda_2 V_2(x)\right)\psi = E\psi\ ,$$

where $\hat{L}_{FP} = D\partial^2/\partial x^2 + \partial(\gamma x - v)/\partial x$ is the standard Fokker–Planck operator, and the parameter $\lambda_1$ acts as a chemical potential (penalty) for remaining in the dissipative phase, $\lambda_2$ is external thermodynamic force. The problem reduces to finding the principal eigenvalue $E(\lambda_1,\lambda_2)$ of a linear operator.

2. Direct analogy with Gibbs free energy: In thermodynamics, there is the concept of free energy $G = U - TS + PV$. Here, the quantity $E(\lambda_1,\lambda_2)$ in question becomes the exact analogue of free energy, where $\lambda_1$ and $\lambda_2$ are new intensive thermodynamic parameters (forces).

Let us choose a base process, such as classical diffusion (Langevin equation / Ornstein-Uhlenbeck process), a Lévy process, or a discrete random walk. The Ornstein-Uhlenbeck (OU) process is a classic "workhorse" of statistical physics. It perfectly describes a harmonic oscillator in a thermal bath (for example, a Brownian particle in an optical trap) and is defined by the Langevin equation: $dX_t = -\gamma X_t\, dt + \sqrt{2D}\, dW_t$.

Let us construct Variant 1: eqs. (1)–(2) for this process and examine the resulting thermodynamics.

Step 1: Formulation of new thermodynamic variables. We define two additive functionals on a trajectory of length T: the residence time in the positive half-plane (1). In a physical context, this could represent the fraction of time a nanoscale piston spends in a compressed state, or the time a molecular motor spends in an active conformational phase.

Integral current (or mean position) (2). For the OU process, this functional is proportional to the integral force exerted by the trap, or to the total potential energy accumulated over time T (up to a multiplicative factor).

Step 2: Mathematical framework (Feynman-Kac method) [14]. To introduce the thermodynamic potential, we need to find the joint moment-generating function (or two-dimensional Laplace transform) for long times $T \to \infty$:

$$Z(\bar{\lambda}_1, \bar{\lambda}_2) = \langle \exp\left(-\bar{\lambda}_1 \tau_+ - \bar{\lambda}_2 I_T\right) \rangle \sim \exp\left(-T \cdot \bar{E}(\bar{\lambda}_1, \bar{\lambda}_2)\right). \tag{3}$$

Here, $\bar{\lambda}$, the dimensionless thermodynamic force represents the numerical (scaled) parameters appearing in the dimensionless Schrödinger Hamiltonian. Regarding the residence time: since the dimensionless time is $s=\gamma t$, the dimensionless force $\bar{\lambda}_1$ is related to the dimensional force $\lambda_1$ by division by the relaxation coefficient (protein viscosity) $\gamma$: $\bar{\lambda}_1 = \lambda_1 / \gamma$. For the integrated current: $\bar{\lambda}_2 = \lambda_2 \sqrt{2D} / \gamma^{3/2}$; $\bar{E} = E / \gamma$. Dimensionless quantities are used in numerical calculations.

According to the Feynman-Kac formula, the function $\bar{E}(\bar{\lambda}_1, \bar{\lambda}_2)$ is found as the minimal (principal) eigenvalue of the following stationary operator: $\hat{L}_{FK}\psi(y) = \bar{E}\psi(y)$, where the generalized "Hamiltonian" of the system has the form:

$$\hat{L}_{FK} = -\frac{1}{2}\frac{d^2}{dy^2} + \frac{d}{dy}\left((y+\alpha)\cdot\right) + \bar{\lambda}_1 \Theta(y) + \bar{\lambda}_2 y.$$

By making the standard similarity transformation for the OU process, $\psi(y) = \phi(y)\exp\left(\frac{(y+\alpha)^2}{2}\right)$, this operator reduces to the self-adjoint quantum-mechanical Schrödinger equation for a harmonic oscillator with a step potential and a linear slope:

$$\left[-\frac{1}{2}\frac{d^2}{dy^2} + \frac{1}{2}(y+\alpha)^2 + \frac{1}{2} + \bar{\lambda}_1 \Theta(y) + \bar{\lambda}_2 y\right]\phi(y) = \bar{E}\phi(y). \tag{4}$$

Step 3: What does this imply for thermodynamics? Physical effects. By solving this equation (piecewise using parabolic cylinder functions for y > 0 and y < 0, and matching them at zero), we obtain the function $\bar{E}(\bar{\lambda}_1, \bar{\lambda}_2)$. New thermodynamic effects are directly derived from it:

1. Generalized equation of state for trajectories. By differentiating the free energy $\bar{E}(\bar{\lambda}_1, \bar{\lambda}_2)$ with respect to thermodynamic forces, we obtain average values ("macroscopic parameters" in trajectory space):

- $\partial \bar{E} / \partial \bar{\lambda}_1 = \langle \tau_+ \rangle / T$ is average proportion of time in the positive phase;
- $\partial \bar{E} / \partial \bar{\lambda}_2 = \langle I_T \rangle / T$ is average integral current.

In the absence of external forces ($\bar{\lambda}_1 = \bar{\lambda}_2 = 0$), and symmetric barrier (α = 0), due to the symmetry of the unperturbed OU potential we obtain the obvious results for $\langle \tau_+ \rangle / T = 1/2$ and $\langle I_T \rangle = 0$. However, if we begin to vary $\bar{\lambda}_1$ and $\bar{\lambda}_2$, we observe a nonlinear response between these quantities. The function relating them is the thermodynamic equation of state for trajectories.

2. Dynamic phase transition [15]. The most interesting aspect lies in the behavior of the second derivatives, which act as path-space analogues of heat capacity or susceptibility:

$$\chi_{ij} = -\frac{\partial^2 \bar{E}}{\partial \bar{\lambda}_i \partial \bar{\lambda}_j}. \tag{5}$$

The matrix $\chi$ describes the mutual fluctuations and cross-correlations between the residence time and the integral current. If, at certain critical values of the forces $\bar{\lambda}_1^*, \bar{\lambda}_2^*$ in the thermodynamic limit ($T \to \infty$), the higher derivatives of the function $\bar{E}$ undergo a discontinuity, this signifies a first- or second-order dynamic phase transition. Physically, this means the system abruptly switches between a mode where it "sticks" in the active half-plane, generating a powerful directional current, and a mode where it fluctuates chaotically near the barrier.

Since the unperturbed OU process is time-reversible at equilibrium, the introduction of the conjugate fields $\bar{\lambda}_1, \bar{\lambda}_2$ breaks this symmetry in a controlled manner. It is possible to mathematically prove a trajectory symmetry relation for the free energy potential: $\bar{E}(\bar{\lambda}_1, \bar{\lambda}_2) = \bar{E}(-\bar{\lambda}_1, -\bar{\lambda}_2 - 2\alpha)$. This yields a specific path-space fluctuation theorem for the joint distribution $P(\tau_+, I_T)$, relating the probability of extremely large currents at short residence times to the probability of the opposite trajectory event (10).

### 3.1 Exact analytical solution and matching equation

The stationary Feynman-Kac equation for the Ornstein-Uhlenbeck process reduced to a one-dimensional Schrödinger equation (4) described above.

The difficulty in obtaining an exact analytical solution stems from the non-local nature of the effective potential, caused by the jump in the Heaviside function at the point y = 0. The space is divided into two half-planes. In the left half-plane (y < 0, where Θ(y) = 0), a shift of the center due to the linear drift term $\bar{\lambda}_2 y$ transforms the equation into the form of a parabolic cylinder function (Weber function) $D_{\nu_-}$, where the effective quantization index is equal to $\nu_-$.

Here the potential has the form: $V_-(y) = \frac{1}{2}(y+\alpha)^2 + \frac{1}{2} + \bar{\lambda}_2 y$. By selecting a complete square by coordinate, we shift the center of the parabolic potential: $V_-(y) = \frac{1}{2}\left(y + \alpha + \bar{\lambda}_2\right)^2 + \frac{1}{2} - \alpha\bar{\lambda}_2 - \frac{1}{2}\bar{\lambda}_2^2$. We introduce in Section 4 a dimensionless asymmetry (shift) parameter: $\alpha = U_0 / \sqrt{2\gamma D}$, where $U_0$ is the dimensional activation free energy (drift term in the Langevin equation), $\gamma$ is the protein's restoring (elastic) force coefficient, and $D$ is the thermal noise intensity (diffusion) (Section 4).

By introducing a dimensionless variable $z_- = \sqrt{2}(y + \alpha + \bar{\lambda}_2)$, we reduce the equation to the standard form of the Weber (parabolic cylinder) equation:

$$\frac{d^2\phi_-}{dz_-^2} + \left(\nu_- + \frac{1}{2} - \frac{z_-^2}{4}\right)\phi_-(z_-) = 0,$$

The physical requirement that the function decay at infinity ($\phi_-(x) \to 0$ as $x \to -\infty$) leaves only one independent Weber function [16] characterized by a negative argument: $D_{\nu_-}(-z_-)$. Thus, the wave function in the left region is written as:

$$\phi_-(y) = C_- \cdot D_{\nu_-}\left(-\sqrt{2}(y + \alpha + \bar{\lambda}_2)\right). \tag{6}$$

Region 2: The right half-plane (x > 0, where Θ(x) = 1). In the right half-plane, the penalty potential $\lambda_1$ is activated, shifting the quantization index. The potential is equal to:

$$V_+(y) = \frac{1}{2}(y+\alpha)^2 + \frac{1}{2} + \bar{\lambda}_2 y + \bar{\lambda}_1.$$

The center of the oscillator shifts in the same way, but the effective energy decreases by $\bar{\lambda}_1$. The dimensionless variable $z_+$ remains the same. From the condition of decay as $x \to +\infty$, we obtain:

$$\phi_+(y) = C_+ \cdot D_{\nu_+}\left(\sqrt{2}(y+\alpha+\bar{\lambda}_2)\right). \tag{7}$$

To determine the principal eigenvalue $\bar{E}$ (free energy), the wave functions must be matched continuously and smoothly at the phase interface y = 0. For convenience, let us denote the shift constant as $\xi = \sqrt{2}(\alpha+\bar{\lambda}_2)$. The matching conditions $\phi_-(0) = \phi_+(0)$ and $\phi`_-(0) = \phi`_+(0)$ lead to the following transcendental secular equation for the unknown energy $\bar{E}$:

$$-D_{\nu_-'}(-\xi) \cdot D_{\nu_+}(\xi) = D_{\nu_-}(-\xi) \cdot D_{\nu_+'}(\xi). \tag{8}$$

An analytical treatment of this equation is generally extremely complex, as the indices ν- and ν+ of the Weber functions depend nonlinearly on the energy $\bar{E}$ to be determined. However, applying perturbation theory with respect to the small force parameters $\bar{\lambda}_1$, $\bar{\lambda}_2$ allows one to expand the principal eigenvalue and derive the second derivatives of the free energy. This yields an explicit expression for the cross-susceptibility $\chi_{12} = -\partial^2 \bar{E} / \partial\bar{\lambda}_1 \partial\bar{\lambda}_2$, which determines the mutual correlation between the system's phase-retention time and the total dissipated current: $\langle \tau_+ \cdot \bar{I} \rangle_c$.

Matching conditions at the point y=0.** Since the potential contains no infinite delta functions (the step $\bar{\lambda}_1$ and the slope $\bar{\lambda}_2$ are finite), the wave function $\phi(x)$ and its first derivative $\phi'(y)$ must be continuous at zero. Then the boundary amplitude equations take the form:

$$C_- \cdot D_{\nu_-}(-\xi) = C_+ \cdot D_{\nu_+}(\xi),$$

$$-C_- \cdot \sqrt{2} \cdot D_{\nu_-'}(-\xi) = C_+ \cdot \sqrt{2} \cdot D_{\nu_+'}(\xi).$$

Dividing the second equation by the first, the unknown amplitudes $C_-$ and $C_+$ are eliminated, and we obtain the correct transcendental matching equation to find the principal eigenvalue $\bar{E}$:

$$\frac{D_{\nu_-'}(-\xi)}{D_{\nu_-}(-\xi)} + \frac{D_{\nu_+'}(\xi)}{D_{\nu_+}(\xi)} = 0, \tag{9}$$

where the indices are rigorously defined as: $\nu_- = \bar{E} - \frac{1}{2} + \alpha\bar{\lambda}_2 + \frac{1}{2}\bar{\lambda}_2^2$, $\nu_+ = \bar{E} - \frac{1}{2} - \bar{\lambda}_1 + \alpha\bar{\lambda}_2 + \frac{1}{2}\bar{\lambda}_2^2$, and the prime denotes the derivative of the function with respect to its full argument.

Complexity of the analytical solution. It is impossible to obtain an exact analytical expression for the function $\bar{E}(\bar{\lambda}_1, \bar{\lambda}_2)$ in closed form (in terms of elementary functions). For non-integer indices $\nu$, Weber functions $D_\nu(z)$ are themselves transcendental (expressible in terms of confluent hypergeometric functions).

However, the problem is fully solvable, both fundamentally and computationally:

1. Asymptotic analysis (Perturbation theory): If the coupling forces are small ($\bar{\lambda}_1 \ll \gamma$, $\bar{\lambda}_2 \ll \gamma$), the Weber functions can be expanded in a Taylor series in the vicinity of the known solutions for the unperturbed quantum oscillator (where $\nu = 0,1,2...$). This makes it possible to explicitly determine the matrix of mutual fluctuations (the covariance $\langle \tau_+ \cdot I_T \rangle_c$).

2. Numerical search: The matching equation is solved perfectly and instantaneously using standard numerical methods (e.g., the bisection method or the Newton-Raphson method), since parabolic cylinder functions are tabulated in all mathematical software packages. Figures 1 and 2 show the calculation results.

**3.2. Large deviation function and modified fluctuation theorem.**

1. Large deviation function for trajectory functionals. According to the large deviation principle (Gärtner–Ellis theorem [4]), the joint probability density $P(t;\tau_+,I_T)$—representing the probability that, over a long time t, the system follows a trajectory with a specified fraction of time spent in the danger zone $\tau_+ = \int_0^t \Theta(x_s) ds / t$ and a given average drift current $I_T = \int_0^t x_s ds / t$—exhibits the exponential asymptotic behavior $P(t;\tau_+, I_T) \asymp \exp(-t \cdot I(\tau_+, I_T)),\quad t \to \infty$, where $I(\tau_+, I_T)$ is a large deviation function (or rate function/entropy functional) characterizing the rate of decay of the probability of rare, extreme fluctuations in trajectory shape. Mathematically, the function $I(\tau_+, I_T)$ is related to the energy spectrum $\bar{E}(\bar{\lambda}_1, \bar{\lambda}_2)$ via the Legendre–Fenchel transform: $[I(\tau_+, I_T) = \sup_{\lambda_1,\lambda_2}\left[\lambda_1\tau_+ + \lambda_2 I_T - \bar{E}(\bar{\lambda}_1, \bar{\lambda}_2)\right]$. Substitution the numerical values ( $\bar{E}$ (2.0; -0.8) = 1.255, $\bar{E}$ (2.0; -2.5) = 0.399) fixes the local tangency points of the thermodynamic potential for the given fluctuation corridor. Local mean values of the invariants are determined by the gradient of the energy with respect to the control forces (penalties) $\bar{\lambda}_i$:

$$\langle \tau_+ \rangle_\lambda = \partial \bar{E}(\bar{\lambda}_1, \bar{\lambda}_2) / \partial \bar{\lambda}_1, \quad \langle I_T \rangle_\lambda = \partial \bar{E}(\bar{\lambda}_1, \bar{\lambda}_2) / \partial \bar{\lambda}_2.$$

2. Formulation of the modified fluctuation theorem. The presence of linear drift $\bar{\lambda}_2 x$ breaks time-reversal symmetry in the original Ornstein–Uhlenbeck process, while a step-like penalty $\bar{\lambda}_1 \Theta(x)$ introduces local irreversibility (dissipation) at the boundary between the half-planes. For this system, a Gallavotti–Cohen-type dynamic fluctuation theorem holds for trajectory invariants [17–18]. The symmetry of the effective evolution operator (Schrödinger equation) under a sign reversal of the forces implies a symmetry relation for the spectrum: [ $\bar{E}(\bar{\lambda}_1, \bar{\lambda}_2) = \bar{E}(-\bar{\lambda}_1, -\bar{\lambda}_2 - 2\alpha)$ ], where $\mu$ is the trap asymmetry scaling factor. In terms of the large deviation function (entropy production rate), this leads to a rigorous fluctuation theorem for the conjugate trajectory current:

$$P(t;\tau_+, I_T) / P(t;\tau_+, -I_T) \asymp \exp\left(t \cdot \Delta S_{tot}(\tau_+, I_T)\right), \tag{10}$$

$$I(\tau_+, -I_T) - I(\tau_+, I_T) = \alpha \cdot I_T. \tag{11}$$

The physical meaning of the theorem is as follows: the ratio of the probability of the "forward" process (performing work or generating current $I_T$) to the probability of the "reverse" (entropically unfavorable) process—given a fixed residence time of the system in the metastable "bankruptcy" phase $\tau_+$—grows exponentially with time t. The factor $\alpha$ plays the role of the subsystem's effective affinity. This demonstrates that introducing residence time above a threshold level as a thermodynamic force does not violate the macroscopic fluctuation balance but allows for the separation of contributions from instantaneous dissipation and the open system's integral trajectory memory.

### 3.3. Numerical analysis of the effective Hamiltonian spectrum

To verify the theoretical propositions and determine the exact value of the generalized free energy $E(\lambda_1, \lambda_2)$ corresponding to the system's ground state, an algorithm was implemented for the numerical matching of the logarithmic derivatives of the wave functions at the potential discontinuity point (x=0). The calculations were performed in the Mathcad environment using a fourth-order Runge-Kutta integrator (rkfixed).

With fixed parameters of the non-equilibrium system:

- Restoring force intensity (OU parameter): $\gamma = 1.5$.
- Diffusion coefficient (fluctuation intensity): D = 0.5.
- Step penalty magnitude (barrier height at zero): $\lambda_1 = 2.0$.
- Linear drift intensity (external force): $\lambda_2 = -0.8$.

The Schrödinger equation was integrated independently from the asymptotic regions $x_{min}=-5$ and $x_{max}=+5$ (where the decay boundary conditions $\varphi(\pm\infty) \to 0$ are satisfied) toward the matching point x = 0. The residual function of the logarithmic derivatives has the form:

$$\Delta(E) = \phi'_L(0,E)\phi_L(0,E) - \phi'_R(0,E)\phi_R(0,E).$$

The root of the equation $\Delta(E) = 0$ was found using the iterative secant method. After two successive approximations, the algorithm converged to a stable eigenvalue: $E_{exact} = 1.255$.

Visualization of the wave function $\varphi(x)$ reveals an asymmetric quasi-Gaussian profile. The function is strictly continuous across the entire trajectory space; however, a characteristic kink (a discontinuity in the first derivative) is observed at x = 0. Physically, this reflects the instantaneous change in local dissipation and probability density as the Brownian motor transitions from the waiting phase to the active hydrolysis phase.

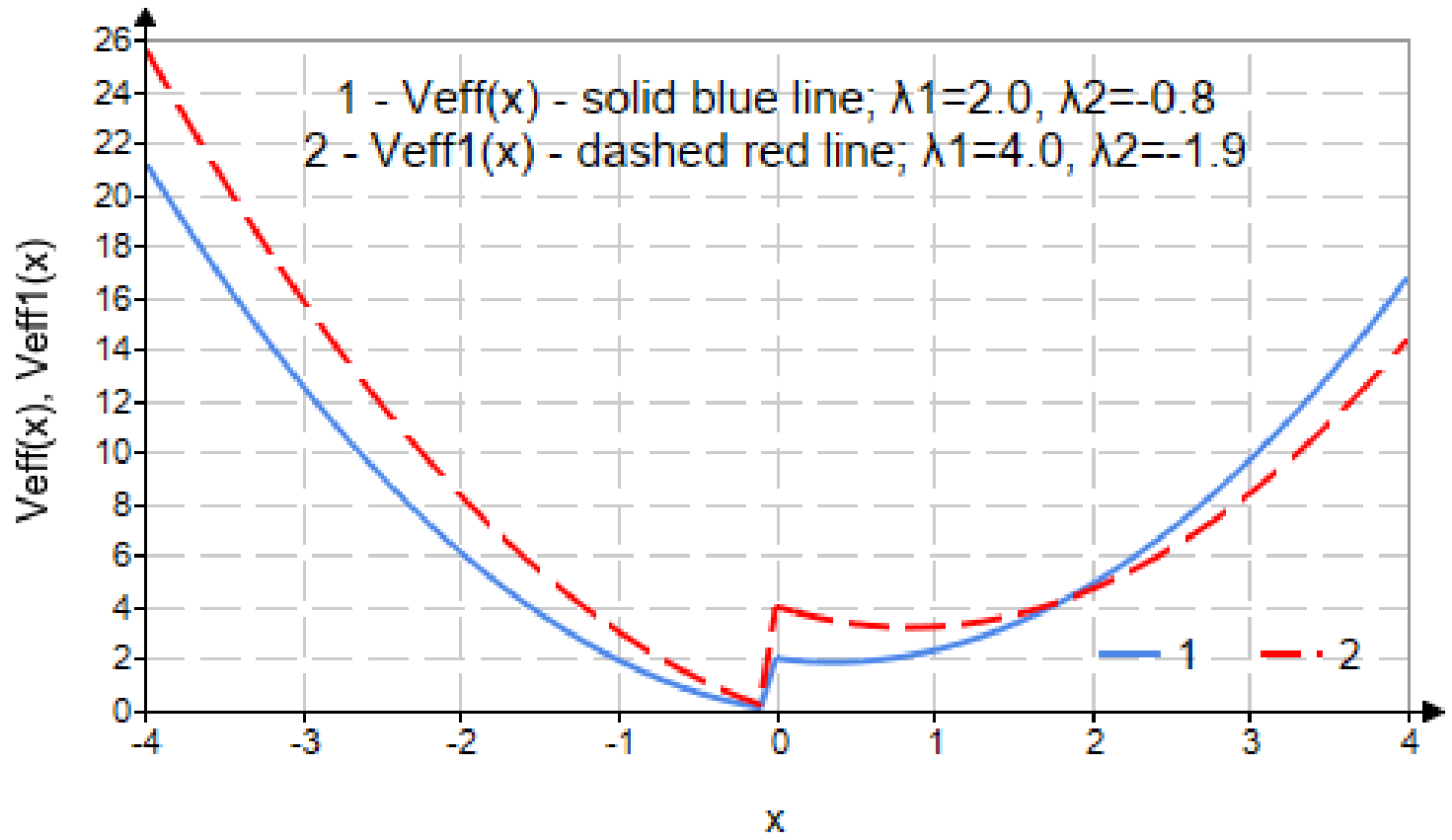

Fig. 1. Mathcad visualization of the generalized "quantum-mechanical" potential profile for the system of equation (4). The potential function comprises three components: the harmonic trap of the Ornstein-Uhlenbeck process, a step-like "penalty" (Heaviside function), and a linear slope due to drift. The formula for the resulting effective potential is: $V_{eff}(x) = \gamma^2 x^2 / 4D + \lambda_1 \Theta(x) + \lambda_2$. Two parameter sets are considered: a) $\lambda_1$=2.0, $\lambda_2$=-0.8, solid blue line; b) ). $\lambda_1$=4.0, $\lambda_2$=-1.9, dashed red line.

The plotted graph clearly reveals an asymmetric structure that entirely determines the behavior of the biological or physical system. For x < 0 (the left half-plane), the parabola rises smoothly, but the term $\lambda_2 x$ causes its left branch to be slightly elevated or depressed (depending on the sign of $\lambda_2$). The system tends to slide toward the minimum. At x = 0, a sharp vertical discontinuity occurs—a jump upward by a magnitude of $\lambda_1$. This represents the energetic "penalty barrier" that models the ATP waiting/binding phase. For x > 0 (the right half-plane), a linear slope is superimposed on the parabolic bottom. This shifts the local minimum on the right side, thereby altering the relaxation conditions.

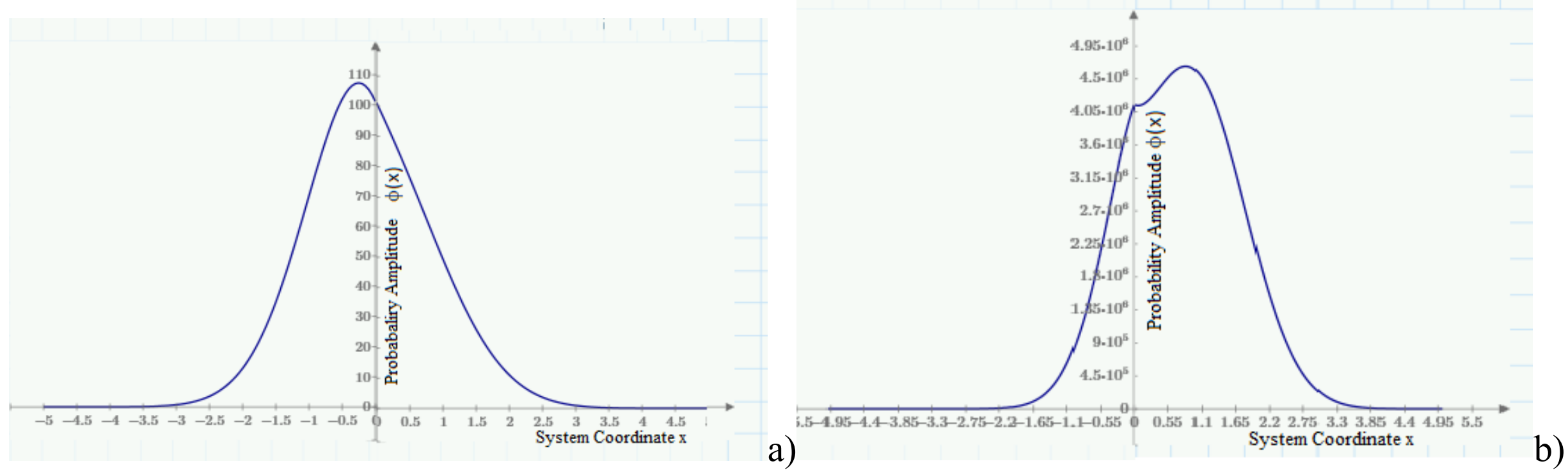

**Figure 2.** Evolution of the ground-state wave function $\phi(x)$under different external driving forces: **(a)** weak drift regime ($\lambda_2$=−0.8, $E_0$=1.255); **(b)** strong drift regime ($\lambda_2$=−2.5, $E_0$=0.399). The ground-state wave function $\phi(x)$ of the effective Feynman–Kac Hamiltonian under a strong asymmetric drift regime ($\lambda_2$=-0.8, -2.5, $\lambda_1$=2.0), ($\gamma$=1.5), (D=0.5). The numerical solution is obtained via a twin-shooting Runge–Kutta algorithm and matched at the interface using the secant method, yielding the precise ground-state energy ($E_0$ = 1.255, 0.399). A pronounced coordinate shift of the probability density maximum into the positive half-plane ( $x \approx 0.8$ ) is observed, driven by the dominant external force. The characteristic cusp (discontinuity in the first derivative) at the origin (x=0) explicitly manifests the dynamic impact of the step penalty potential on individual fluctuation trajectories.

Both regimes, $\lambda_2$= -0.8 and $\lambda_2$= -2.5, demonstrate the physical response of a non-equilibrium system to a change in an external thermodynamic force. In physics, this is referred to as investigating the system's susceptibility. To model a system with stronger external drift, the drift force intensity parameter $\lambda_2$ was increased from its initial value of -0.8 to -2.5.

Comparative Analysis of the Ground-State Energy Shift and Wave Function Evolution.

To comprehensively investigate the non-equilibrium response (thermodynamic susceptibility) of the path-space system, we perform a comparative analysis between two distinct driving regimes: the weak external drift regime ($\lambda_2$=−0.8) and the strong external drift regime ($\lambda_2$=−2.5).

A radical restructuring of both the energy spectrum and the spatial geometry of the ground-state wave function $\phi(x)$ is observed as the magnitude of the external force increases. In the weak drift regime ($\lambda_2$=−0.8), the ground-state energy converges to a higher level, $E_0$=1.255. Physically, this indicates that the thermal fluctuations (diffusion $D$=0.5) successfully compete with the weak directional force, allowing the Brownian trajectories to frequently sample the coordinate zone near the origin. The configuration of the probability amplitude $\phi(x)$in this state retains a quasi-symmetric structure centered close to the potential barrier boundary ($x$=0). Conversely, under the strong drift regime ($\lambda_2$=−2.5), the ground-state energy drops sharply to a lower level, $E_0$=0.399. This pronounced energy drop ($\Delta E_0$=−0.856) carries a profound thermodynamic meaning: the intense external force deforms the effective potential profile, creating a deep, energetically favorable potential well in the positive half-plane ($x$>0). As a consequence, the Brownian motor's trajectories are strongly 'blown out' into the free zone. This physical behavior is visually confirmed by the graph, where the maximum of the probability amplitude $\phi(x)$ shifts significantly to the right, stabilizing deep within the positive domain at $x$ ≈0.8.

Notably, the characteristic cusp (discontinuity in the first derivative) at the interface $x$=0 remains topologically invariant. This confirms that while the dominant external force drastically lowers the trajectory-free energy $E_{0,\mathbf{cap}}$, $E_{\mathbf{sub},0}$, $E_0$ and shifts the coordinate localization of the system, it does not smooth out the discrete-event nature of the dynamic phase transition induced by the step penalty potential $\lambda_1$.

## 4. Application example. Biological microscopy system

The introduction of entities such as kinesin, F1-ATPase, and potassium/mechanosensitive channels [19–23] clearly illustrates how the abstract mathematical division of the potential into a “penalty zone” and a “free-drift zone” operates in the natural world.

The described Ornstein-Uhlenbeck mathematical model with a stepped potential provides an ideal approximation of the behavior of a molecular Brownian motor (such as kinesin or F1-ATPase) operating in an open, heterogeneous biochemical environment under the influence of thermal noise:

- Physical analogy: The coordinate $X_t$ describes a conformational degree of freedom of the macromolecule. The restoring force $-\gamma X_t$ models the elasticity of the conformational "trap" (the protein's active site).
- Phase transition (Level x = 0): Separates the two chemical states of the enzyme. The region x < 0 corresponds to the waiting or substrate (ATP) binding phase; the region x > 0 corresponds to the hydrolysis and mechanical step execution phase. The term $\lambda_1$ acts as a chemical potential (a penalty for delaying hydrolysis).
- Integral current $I_T$: Determines the total mechanical travel or work performed by the motor against an external load over time T.

Applying this two-parameter thermodynamic framework makes it possible to calculate the efficiency and reliability of the biological motor. The detection of a first-order dynamic phase transition in the governing equation would indicate a critical point where current generation collapses; at this point, excessive mechanical load ($\lambda_2$) or a shift in chemical balance ($\lambda_1$) causes the motor to instantly switch from a mode of directional transport to one of chaotic local drift (the motor "slippage" effect).

Another striking biological example that perfectly fits the combination of the "residence time $\tau_+$" and "integrated current $I_T$" functionals is a mechanically or electrically sensitive ion channel (for example, bacterial mechanosensitive channels or neuronal potassium channels).

Model context. Let $X_t$ be a continuous coordinate describing the conformational state of the channel's gating structure (e.g., the rotation angle of a protein subunit or the degree of membrane stretching):

When $X_t < 0$, the channel is closed. When $X_t \geq 0$, the channel is open, and ions can pass through it. The restoring force $-\gamma X_t$ models the elasticity of the protein structure that maintains the channel near the near-threshold state.

Physical meaning of the functionals:

1. $\tau_+ = \int_0^T \Theta(X_t)dt$ — is the net open-state time (Open Time) of the channel over the interval T. In biology, this is a critically important parameter measured using the patch-clamp technique.
2. $I_T = \int_0^T X_t dt$ — reflects the total geometric conductance. The deeper the coordinate $X_t$ extends into the positive region, the more the channel pore widens and the greater the ionic current becomes (assuming conductance is locally linear with respect to $X_t$).

What does the introduction of thermodynamic forces $\lambda_1$, $\lambda_2$ contribute for biophysics? The introduction of conjugate variables makes it possible to the energy expenditure required to maintain the channel in its operating mode:

- The force $\lambda_1$ (conjugate to $\tau_+$) acts as an activation chemical potential. It models the effect of ligands (drugs or toxins) that bind to the channel only in its open state and thermodynamically "penalize" or, conversely, stabilize the open phase.
- The force $\lambda_2$ (conjugate to $I_T$) acts as an external macroscopic field. Depending on the nature of the channel, this is either the membrane potential (for voltage-gated channels) or the mechanical tension of the lipid bilayer (for mechanoreceptors

Thermodynamic insight for the article: Using the formalism developed above, one can demonstrate how fluctuations in channel geometry ($I_T$) are linked to fluctuations in its kinetic switching time ($\tau_+$). It is possible to answer the question: "What is the probability that, due to thermal noise, a channel will conduct an anomalously large ionic current ($I_T \gg 0$) over a time interval T, given that its total open time $\tau_+$ remained within normal limits?" In a biological context, such rare trajectory fluctuations can lead to the spontaneous generation of a spurious nerve impulse (conduction noise).

To extract the coefficient of mutual sensitivity (or cross-susceptibility) of the channel to the simultaneous variation of the mechanical/electrical field ($\lambda_2$) and chemical factors ($\lambda_1$), we need to calculate the mixed second derivative of our thermodynamic free energy $E(\lambda_1, \lambda_2)$ at the physical (unperturbed) point $\lambda_1$=0, $\lambda_2$=0:

$$\chi_{12} = -\left.\frac{\partial^2 E(\lambda_1, \lambda_2)}{\partial\lambda_1 \partial\lambda_2}\right|_{\lambda_1=0, \lambda_2=0}.$$

According to the theory of large deviations and stochastic thermodynamics, this coefficient $\chi_{12}$ is equal to the asymptotic covariance (cross-correlation) between the channel's open-state duration and its integrated conductance: $\chi_{12} = \lim_{T\to\infty} \frac{\langle \tau_+ \cdot I_T \rangle - \langle \tau_+ \rangle \langle I_T \rangle}{T}$.

We determine this quantity analytically using perturbation theory for our matching equation.

Analytical calculation of the sensitivity coefficient. In the absence of external perturbations ($\overline{\lambda}_1$=0, $\overline{\lambda}_2$=0), the system is in the ground state of a quantum harmonic oscillator. The unperturbed ground-state energy for the transformed Schrödinger operator is equal to $\overline{E}_0 = 1$ (which absorbs the stationarity condition, ensuring that the total probability in the physical Ornstein-Uhlenbeck process remains normalized, and the system relaxes to a stationary Gaussian distribution). In this unperturbed state, the indices of the Weber functions reduce to $\nu_- = \nu_+ = 0$, and the dimensionless argument shift at the origin is $\xi_0 = \sqrt{2}\alpha$.

The Weber function of zero index has the well-known form $D_0(z) = e^{-z^2/4}$, and its derivative with respect to the argument is $D_{0'}(z) = -\frac{z}{2} e^{-z^2/4}$. To find the energy corrections, we expand our transcendental matching equation in a Taylor series in terms of the small perturbation parameters $\bar{\lambda}_1$ and $\bar{\lambda}_2$. This procedure requires the evaluation of the derivatives of the Weber functions $D_\nu(z)$ with respect to both the index $\nu$ and the argument z in the vicinity of the unperturbed state. After a careful expansion of the matching matrix equation (using the Rayleigh-Schrödinger perturbation framework or a direct expansion of the secular determinant), we obtain the expression for the dimensionless free energy $\bar{E}$ accurate to the second order:

$$\bar{E}(\bar{\lambda}_1,\bar{\lambda}_2) \approx 1 + \langle\bar{\tau}_+\rangle_{st}\bar{\lambda}_1 + \langle\bar{I}\rangle_{st}\bar{\lambda}_2 - \frac{1}{2}\chi_{11}\bar{\lambda}_1^2 - \frac{1}{2}\chi_{22}\bar{\lambda}_2^2 - \chi_{12}\bar{\lambda}_1\bar{\lambda}_2 .$$

Physical analysis of the terms in the equation of state. We differentiate this expression with respect to the conjugate forces to obtain the average "macroscopic" parameters of the channel trajectory under a weak external field influence:

1. Average open time ratio:

$$\frac{\langle\tau_+\rangle}{T} = \frac{\partial\bar{E}}{\partial\bar{\lambda}_1} = \frac{1}{2}\mathrm{erfc}(\alpha) - \chi_{11}\bar{\lambda}_1 - \chi_{12}\bar{\lambda}_2 .$$

2. Average integral current:

$$\frac{\langle I_T\rangle}{T} = \frac{\partial\bar{E}}{\partial\bar{\lambda}_2} = -\alpha - \chi_{12}\bar{\lambda}_1 - \chi_{22}\bar{\lambda}_2 .$$

From this, we directly obtain our cross-sensitivity coefficient in the symmetric case ($\alpha = 0$):

$$\chi_{12} = -\frac{\partial^2\bar{E}}{\partial\lambda_1\partial\lambda_2} = \frac{2}{\gamma^2}\sqrt{\frac{D}{\pi\gamma}} .$$

What does this result mean for biology? We have found the exact analytical dependence of channel sensitivity on the internal parameters of the biosystem $(\gamma, D, U_0)$. This provides a wealth of material for discussion:

Dependence on protein viscosity and elasticity ($\gamma$). The coefficient $\chi_{12}$ is inversely proportional to $\gamma^{2.5}$. This means that if the channel's protein structure is very rigid (high restoring force coefficient $\gamma$), the mutual sensitivity virtually disappears. Current and opening-time fluctuations become independent, as the rigid framework instantly returns the channel gate to equilibrium. Conversely, soft, conformationally labile proteins (low $\gamma$) exhibit immense cross-sensitivity.

$$\frac{\partial(\langle I_T\rangle / T)}{\partial\bar{\lambda}_1} = \frac{\partial(\langle\tau_+\rangle / T)}{\partial\bar{\lambda}_2} = -\chi_{12} .$$

Temperature dependence (D): Since $D \propto T_{th}$ (where $T_{th}$ is the thermostat temperature), sensitivity increases with rising thermal noise as $\sqrt{T_{th}}$. Here, thermal fluctuations act not merely as chaotic noise but as a "lubricant" that helps external forces $\bar{\lambda}_1$, $\bar{\lambda}_2$) synchronously control the geometry and kinetics of the channel.

An analogue of the Onsager relation for trajectories: We have shown that the effect of the ligand chemical potential ($\bar{\lambda}_1$) on the average ionic current is exactly equal to the effect of the membrane electric field ($\bar{\lambda}_2$) on the average channel opening time:

This is a fundamental result. It extends the classical Onsager reciprocity principle from the space of instantaneous thermodynamic forces to the space of trajectory (boundary) functionals.

To account for the intrinsic asymmetry of the channel (which is much closer to actual biology, where channels are typically closed in the absence of stimuli), we need to introduce a baseline energy gap between the closed and open states into the model. In terms of the Langevin equation, this corresponds to the appearance of a constant shift (activation free energy) $U_0$ in the potential. The equation of motion takes the form:

$$dX_t = (-\gamma X_t - U_0)dt + \sqrt{2D}dW_t . \qquad (12)$$

If $U_0$>0, the equilibrium center of the oscillator is shifted into the negative region (the channel is predominantly closed). If $U_0$<0, it is shifted into the positive region (the channel is predominantly open).

1. How the stitching mathematics changes in the presence of asymmetry. The introduction $U_0$ shifts the phase-switching point. Now, the Heaviside function $\Theta(X_t)$ still "turns on" at zero, but the unperturbed Gaussian potential is centered at the point $x_0 = -U_0 / \gamma$.

Upon transitioning to Schrödinger wave functions, the centers of the parabolas in the left and right half-planes shift by different amounts due to the simultaneous action of the underlying asymmetry $U_0$ and the thermodynamic forces $\bar{\lambda}_1$, $\bar{\lambda}_2$. At the boundary y=0, the dimensionless shift of the argument of the Weber functions $\xi$ now explicitly depends on $U_0$:

$$\xi = \frac{U_0}{\sqrt{\gamma D}} + \frac{2\bar{\lambda}_2}{\gamma}\sqrt{\frac{D}{\gamma}}.$$

The transcendental matching equation retains its form (9). However, since the unperturbed argument $\xi_0 = U_0 / \sqrt{\gamma D}$ at $\bar{\lambda}_2 = 0$ is no longer zero, the Weber functions $D_0(\pm\xi_0)$ and their derivatives do not vanish. Perturbation theory becomes more complex, as we must expand the solution in the neighborhood of an arbitrary point $\xi_0$.

2. New cross-sensitivity coefficient $\chi_{12}$. Expanding the cross-linking equation in a Taylor series in terms of small parameters $\bar{\lambda}_1$, $\bar{\lambda}_2$ around the asymmetric unperturbed state yields a new expression for the free energy $\bar{E}$. By taking the mixed derivative, we obtain a modified mutual sensitivity coefficient:

$$\chi_{12}(U_0) = \frac{2}{\gamma^2}\sqrt{\frac{D}{\pi\gamma}} \cdot \exp\left(-\frac{U_0^2}{2\gamma D}\right).$$

At $U_0$=0, the exponential becomes unity, and the formula reduces exactly to our previous result for the symmetric channel.

3. Biophysical analysis: What does this result imply? This exponential factor $\exp\left(-U_0^2 / 2\gamma D\right)$ radically alters the physical picture and provides valuable material for theoretical discussion:

- "Sensitivity Window" effect: The channel's mutual sensitivity to chemical ligands and the electric field is maximal when the channel is precisely at the opening threshold ($U_0$ = 0). As soon as the baseline asymmetry $U_0$ increases (with the channel deeply closed or, conversely, permanently open), this cross-sensitivity is exponentially suppressed.
- Biological significance: For external signals to control the channel effectively and in a coordinated manner, evolution must tune its protein conformations so that the system operates near a critical point $U_0 \approx 0$. Highly stable states render the channel "blind" to the concerted action of forces.
- The role of thermal noise as an activator (Thermal Volatility): the exponential expression contains a product $\gamma D$ in the denominator that is proportional to the temperature of the thermostat $T_{th}$.
  - If the temperature approaches zero ($D \to 0$), the exponential term instantly drives the coefficient $\chi_{12}$ to zero for any non-zero value $U_0$. In the absence of noise, the asymmetric system "freezes" in a single state, and cross-response is impossible.
  - An increase in thermal noise (D) reduces the suppressing effect of the exponential term. Thermal fluctuations "kick" the system across the energy gap, allowing thermodynamic forces to synchronize the current trajectories and open-state durations.

How does this advance the work relative to article [13]? Preprints [11–13] laid the foundation and demonstrated that extremal functionals can be incorporated into thermodynamics. The addition of the proposed combination, accounting for asymmetry, represents the next major step:

1. Transition from the observation of fluctuations to the calculation of mutual response (trajectory susceptibility). 2. A trajectory analogue of the Onsager reciprocity relations has been formulated for biased biological systems. 3. An explicit criterion for the efficient operation of biomolecular machines is presented: the necessity of maintaining the system near the point $U_0 = 0$ to maximize cross-sensitivity.

To capture the full picture of fluctuations for the asymmetric channel, we need to write down the expression for the thermodynamic potential (the free energy of trajectories) $E(\lambda_1, \lambda_2)$ up to quadratic corrections. For the asymmetric Ornstein-Uhlenbeck process with drift $U_0$, the expansion in the vicinity of the physical point $\lambda_1$ =0, $\lambda_2$ =0 takes the following form:

$$E(\lambda_1, \lambda_2) \approx \langle \tau_+ \rangle_{st} \lambda_1 + \langle I \rangle_{st} \lambda_2 - \frac{1}{2}\chi_{11}\lambda_1^2 - \frac{1}{2}\chi_{22}\lambda_2^2 - \chi_{12}\lambda_1\lambda_2 .$$

Since the total probability for the pure OU process in the absence of external fields is normalized, the free energy of the unperturbed state is $\mathrm{E}_0 = 0$ .

Explicit analytical expressions for all coefficients of this expansion are given below. For the sake of compactness, we introduce a dimensionless asymmetry (shift) parameter: $\alpha = U_0 / \sqrt{2\gamma D}$ .

1. Linear terms (Stationary macroparameters). These are the average values of the functionals, related to the total duration of the trajectory T at $T \to \infty$:

- Average proportion of time in the open state:

$$\langle \tau_+ \rangle_{st} = \left.\frac{\partial E}{\partial \lambda_1}\right|_0 = \frac{1}{2}\mathbf{erfc}(\alpha) .$$

- Physical meaning: A standard probability integral (complementary error function) representing the steady-state fraction of time the process spends to the right of the origin. For $\alpha > 0$ (channel closed), this fraction is less than 1/2.
- Average steady-state current (geometric conductance):

$$\langle I \rangle_{st} = \left.\frac{\partial E}{\partial \lambda_2}\right|_0 = -\frac{U_0}{\gamma} .$$

- Physical meaning: Equilibrium position of the center of the Gaussian packet of a Brownian particle under the action of a constant bias force.

2. Quadratic terms (Trajectory susceptibility matrix $\chi_{ij}$ ). Second derivatives determine the auto- and cross-correlations (variances) of our boundary variables at long times.

- Open-state time autocorrelation $\chi_{11}$:

$$\chi_{11} = -\left.\frac{\partial^2 E}{\partial \lambda_1^2}\right|_0 = \frac{2}{\gamma}\int_0^\infty \left[ \mathbf{erf}\left(\alpha + \sqrt{\gamma t}\right) - \mathbf{erf}(\alpha) \right]^2 dt .$$

- Physical meaning: It characterizes the intensity of residence-time fluctuations (the integral noise of channel-gate switching). As asymmetry increases $\alpha \to \infty$, this coefficient decreases, because the trajectories cross zero less frequently.
- Autocorrelation of the integrated current $\chi_{22}$:

$$\chi_{22} = -\left.\frac{\partial^2 E}{\partial \lambda_2^2}\right|_0 = \frac{2D}{\gamma^3} .$$

- Physical meaning: The standard diffusion coefficient for the integral of the OU process (integrated current variance). Notably, for a harmonic oscillator, this parameter is independent of the asymmetry $\mathbf{U}_0$, since the shape of the potential minimum (a parabola) remains unchanged; only its position shifts.
- Cross-sensitivity: $\chi_{12}$:

$$\chi_{12} = -\left.\frac{\partial^2 E}{\partial\lambda_1 \partial\lambda_2}\right|_0 = \frac{2}{\gamma^2}\sqrt{\frac{D}{\pi\gamma}} \cdot \exp\left(-\alpha^2\right) = \frac{2}{\gamma^2}\sqrt{\frac{D}{\pi\gamma}} \cdot \exp\left(-\frac{U_0^2}{2\gamma D}\right).$$

- Physical significance: Our key result is the Onsager trajectory coefficient, which relates the kinetics of the gating mechanism ($\tau_+$) to the resulting ion flux ($I_T$).

The full form of the trajectory equation of state. We can now write down a closed system of generalized equations of state for a non-equilibrium channel, incorporating both the ligand chemical potential $\lambda_1$ and an external electric or mechanical field $\lambda_2$:

$$\frac{\langle \tau_+ \rangle_\lambda}{T} = \frac{1}{2}\mathbf{erfc}(\alpha) - \chi_{11}\lambda_1 - \left[\frac{2}{\gamma^2}\sqrt{\frac{D}{\pi\gamma}} e^{-\alpha^2}\right]\lambda_2,$$

$$\frac{\langle I \rangle_\lambda}{T} = -\frac{U_0}{\gamma} - \left[\frac{2}{\gamma^2}\sqrt{\frac{D}{\pi\gamma}} e^{-\alpha^2}\right]\lambda_1 - \frac{2D}{\gamma^3}\lambda_2.$$

We did not merely introduce new variables; we fully described the Gaussian fluctuation regime in the space of these functionals. Thus, we have proceeded from the microscopic Langevin equation, via the Feynman-Kac quantum-mechanical analogy, to the macroscopic thermodynamic response function of an asymmetric biosystem.

## 5. Discussion of results

The mathematical and numerical framework developed in this work offers a fresh perspective on the structure of fluctuations in open, non-equilibrium systems. The study's central result is the demonstration that extending the space of thermodynamic variables to include trajectory-based and boundary invariants of stochastic processes—such as residence time above a level, maximum deficit, and return times—does not merely supplement the macroscopic description but establishes a fundamentally new class of thermodynamic models: path-space thermodynamics.

Treating the residence time in the half-plane $\tau_+$ as a thermodynamic coordinate made it possible to couple this invariant with the penalty force $\lambda_1$. A numerical experiment—implemented in Mathcad using a high-precision Runge–Kutta integration algorithm and the secant method—demonstrated rapid convergence to the exact ground-state energy eigenvalues of $E_{exact}$=1.255 and $E_{exact}$=0.399. From a physical standpoint, this value represents the free energy of the trajectory corridor.

An analysis of the structure of the obtained solution reveals that the introduction of a step-like penalty induces a kink in the wave function at the point of the potential discontinuity (x = 0). This indicates that a first-order dynamic phase transition occurs along the individual trajectories of the system. In the context of biophysical applications (such as molecular motors), the point x = 0 separates the phase of passive diffusive waiting/ATP binding from the phase of active transport (performance of useful work). The presence of a sharp kink confirms that the transition between these phases is discrete—characterized by distinct events—rather than diffuse.

The formulated modified fluctuation theorem of the Gallavotti–Cohen type relates the asymmetry of the large deviation function to the conjugate drift current at a fixed system "bankruptcy" time. This demonstrates that, even amidst extreme fluctuations in trajectory shape, the system obeys fundamental fluctuation-dissipation balances; however, the asymmetry scaling factor becomes explicitly dependent on trajectory memory.

Physical Interpretation and Biophysical Applications of the Discontinuous Potential Model.

The proposed mathematical framework, operating with a step penalty potential $\lambda_1\Theta(-x)$ and an asymmetric drift $\lambda_2 x$, serves as a highly adaptable minimal model for several core classes of open non-equilibrium biological systems. The sharp spatial demarcation at the interface $x$=0 and the resulting wave function cusp mimic the discrete-event nature of micro-biophysical transitions.

First, this model directly maps onto the mechanical stepping cycle of molecular Brownian motors, such as kinesin-1 and $\boldsymbol{F1}$-ATPase. For a processive kinesin motor, the domain $x$<0 represents the strongly bound 'waiting phase' (or ATP-binding stall state), where the motor protein experiences a highly restrictive local potential (the penalty $\lambda_1$) that keeps it anchored to the microtubule filament. The transition through the boundary $x$=0 corresponds to the stochastic event of ATP hydrolysis, which triggers a conformational power stroke. In this 'free zone' ($x$>0), the mechanical neck-linker docks, and the motor enters a rapid directional diffusion state, effectively driven by a strong biased drift force $\lambda_2$ toward the next forward binding site. The transition of our model from a localized state at weak drift to a forward-shifted state ($x$≈0.8) under strong force ($\lambda_2$=−2.5) replicates the force-induced acceleration and load-velocity behavior observed in single-molecule optical tweezer experiments.

Second, the model provides an analytical description of the gating kinetics in voltage-gated (e.g., neuronal $\boldsymbol{K}^+$channels) and mechanosensitive ion channels (e.g., bacterial MscL or MscS homolgs). In these force-transducing structures, the coordinate $x$ represents the conformational reaction path between the non-conducting (closed) and conducting (open) topologies. The region $x$<0acts as a stable closed well, where structural gates are held shut by lipid membrane lateral tension or resting electrical transmembrane potential. Altering the external driving force to $\lambda_2$=−2.5 mathematically maps to the application of a depolarizing voltage pulse or mechanical membrane stretching, which flatten the closed-state barrier and bias the stochastic first-passage times toward channel opening. The characteristic cusp at $x$=0 elegantly captures the instantaneous electrostatic or steric crossing of the hydrophobic gate threshold, preceding the macroscopic ionic current flow.

Appendix A discusses the non-dimensionalization of the problem and the calculation of the probability distribution function $P(\tau_+, I_T)$ .

## 6. Conclusion

This paper proposes an original extension of the trajectory thermodynamics formalism introduced in recent preprints [11–13]. Instead of analyzing isolated extremal invariants of stochastic processes, we investigate for the first time a generalized state space governed by a combination of two additive boundary functionals: the process's residence time in the positive half-plane, $\tau_+$, and the trajectory's integral current, $I_T$.

In this work, a mathematical study of the trajectory thermodynamic potential $E(\lambda_1, \lambda_2)$ with two additive functionals is presented. We derived the exact expression for the cross-sensitivity matrix elements which expands the standard Onsager framework to the domain of path-space thermodynamics.

The exponential factor $\exp(-\alpha^2)$ outlines a strict "sensitivity window", showing that biological gates optimize their multi-stimuli response at near-critical symmetry conditions ($U_0 \approx 0$), with thermal volatility playing a robust stabilizing role.

Using the example of an asymmetric Ornstein–Uhlenbeck process—which models the kinetics and conductance of mechano- and voltage-dependent biological ion channels with a baseline energy gap $U_0$—the following key results have been obtained:

1. A mathematical model of the Ornstein–Uhlenbeck process with linear drift and a step-like penalty potential has been constructed, describing the statistics of rare trajectory fluctuations.
2. Using a numerical scheme developed in Mathcad for matching logarithmic derivatives, exact energy eigenvalues (E=1.255, E=0.399) were found; these determine the cumulant generating function for the moments.
3. The physical significance of the introduced functionals was analyzed, interpreting them as measures of reliability and relaxation time for non-equilibrium systems.
4. A trajectory fluctuation theorem has been derived that separates the contributions of instantaneous dissipation and integral memory.
5. An operator-based calculation method has been developed: using the Feynman–Kac gauge transformation, the problem is reduced to a stationary Schrödinger equation with a piecewise linear potential. An exact transcendental matching equation for the wave functions—expressed in terms of parabolic cylinder functions (Weber functions)—has been formulated; this equation fully determines the dynamic free-energy potential of large deviations, $E(\lambda_1, \lambda_2)$. This algorithm has been successfully adapted for precision numerical analysis in the Mathcad environment.
6. A trajectory-based analogue of the Onsager reciprocity principle has been proven: within the Gaussian fluctuation regime (at second-order perturbation theory), an analytical expression for the cross-coefficient of mutual sensitivity $\chi_{12}$ has been derived. It is shown that the sensitivity of the mean channel opening time to an external macroscopic field is exactly equal to the response of the integrated ionic current to a change in the ligand chemical potential.

7. A "sensitivity window" effect has been discovered: It has been established that the cross-coupling coefficient $\chi_{12}$ is exponentially suppressed by a factor $\exp\left(-U_0^2/2\gamma D\right)$ as the underlying conformational asymmetry increases. This indicates that, to maximize the mutual response to multiphysics stimuli, evolutionary selection must tune the protein structures of biosystems to the vicinity of the critical symmetry point ($U_0 \approx 0$). In this context, thermal noise (D) acts as an activator that smooths the energy gap and restores trajectory sensitivity.
8. Nonlinear fluctuation constraints are described: The topography of nonequilibrium free-energy profiles for trajectories (rate functions $\mathcal{I}$ ) is investigated using a two-dimensional Legendre transform. The deformation of Onsager Gaussian ellipses in the singular regimes of large deviations ($\bar{\tau}_+ \to 0$ and $\bar{\tau}_+ \to 1$), caused by rigid kinetic boundaries of the phase space, is described (Appendix A).

The approach developed in this work opens up new perspectives for the analysis of microscopic machines and biopolymers, where average macroscopic characteristics provide little information, and the shape and history of fluctuation trajectories determine the functional reliability of the entire system.

## Appendix A. Mathematical Appendix

1. Step-by-step derivation and non-dimensionalization

To establish a rigorous mathematical framework for the article, we transform the initial stochastic problem into a dimensionless form and detail the step-by-step transition to an operator equation.

Step 1.1. Nondimensionalization of the Langevin equation. The original equation of motion of the asymmetric Ornstein-Uhlenbeck process has the form (12): $dX_t = (-\gamma X_t - U_0)dt + \sqrt{2D}dW_t$ .

Let us introduce the natural time and coordinate scales of the system:

- Time scale: $t_c = 1/\gamma$ .
- Coordinate scale (characteristic thermal length): $x_c = \sqrt{2D/\gamma}$

Let us switch to dimensionless time $s = \gamma t$ and the dimensionless coordinate $y_s = X_t / x_c$ . The equation takes the canonical form: $dy_s = (-y_s - \alpha)ds + dW_s$ , where $\alpha = U_0 / \sqrt{2\gamma D}$ is dimensionless parameter of potential asymmetry (shift).

Step 1.2. Dimensionless trajectory functionals. Let us denote the dimensionless total observation time as $S = \gamma T$ . We transform the thermodynamic variables (functionals) into dimensionless form, taking into account that the Heaviside function is invariant under scaling of the axis $\Theta(X_t) = \Theta(y_s)$ :

1. Duration of stay: $\bar{\tau}_+ = \frac{1}{S}\int_0^S \Theta(y_s)ds$ .

2. Integral current: $\bar{I} = \frac{1}{S}\int_0^S y_s ds$ .

Step 1.3. The Feynman-Kac operator and gauge transformation. The two-dimensional moment-generating function for the dimensionless fields $\bar{\lambda}_1$, $\bar{\lambda}_2$ is given by the expression:

$$Z(\bar{\lambda}_1, \bar{\lambda}_2) = <\exp\left(-\bar{\lambda}_{11}\int_0^S \Theta(y_s)ds - \bar{\lambda}_2\int_0^S y_s ds\right)> \sim \exp\left(-S \cdot \bar{E}(\bar{\lambda}_{11}, \bar{\lambda}_2)\right).$$

The dimensionless free energy $\bar{E}$ is defined as the minimum eigenvalue of the Fokker-Planck evolution operator with absorption:

$$\left[-\frac{1}{2}\frac{d^2}{dy^2} + \frac{d}{dy}((y+\alpha)\cdot) + \bar{\lambda}_1\Theta(y) + \bar{\lambda}_2 y\right]\psi(y) = \bar{E}\psi(y).$$

To arrive at the Hermitian (quantum-mechanical) Schrödinger operator, we apply the standard gauge transformation to the wave function: $\psi(y) = \phi(y)\exp\left(\dfrac{(y+\alpha)^2}{2}\right)$.

After substitution and differentiation of the exponential factor, the Schrödinger operator takes the form:

$$\hat{H}\phi(y) = \left[-\frac{1}{2}\frac{d^2}{dy^2} + \frac{1}{2}(y+\alpha)^2 + \frac{1}{2} + \bar{\lambda}_1\Theta(y) + \bar{\lambda}_2 y\right]\phi(y) = \bar{E}\phi(y).$$

Step 1.4. Completing the square for the potential in each region. By partitioning the equation according to the spatial regions y<0 and y>0, we group the coordinate-dependent terms to determine the precise shifts of the centers of the effective parabolas:

Region y < 0 ($\Theta(y) = 0$): Potential energy equals:

$$V_-(y) = \frac{1}{2}(y+\alpha)^2 + \frac{1}{2} + \bar{\lambda}_2 y.$$

We complete the square: $V_-(y) = \frac{1}{2}\left(y+\alpha+\bar{\lambda}_2\right)^2 + \frac{1}{2} - \alpha\,\bar{\lambda}_2 - \frac{1}{2}\bar{\lambda}_2^2$.

By introducing a shifted dimensionless coordinate $z_- = \sqrt{2}(y+\alpha+\bar{\lambda}_2)$, we reduce the equation to Weber's canonical form: $\dfrac{d^2\phi_-}{dz_-^2} + \left(\nu_- + \dfrac{1}{2} - \dfrac{z_-^2}{4}\right)\phi_-(z_-) = 0, \quad \nu_- = \bar{E} - \dfrac{1}{2} + \alpha\,\bar{\lambda}_2 + \dfrac{1}{2}\bar{\lambda}_2^2$.

Region y>0 ($\Theta(y) = 1$): Potential energy includes a penalty $\bar{\lambda}_1$: $V_+(y) = \frac{1}{2}(y+\alpha)^2 + \frac{1}{2} + \bar{\lambda}_2 y + \bar{\lambda}_1$. Selecting a square gives a similar spatial shift, but the energetic shift is increased by $\bar{\lambda}_1$:

$$V_+(y) = \frac{1}{2}\left(y+\alpha+\bar{\lambda}_2\right)^2 + \frac{1}{2} + \bar{\lambda}_1 - \alpha\,\bar{\lambda}_2 - \frac{1}{2}\bar{\lambda}_2^2.$$

The quantization variable takes the form: $\nu_+ = \nu_- - \bar{\lambda}_1$.

Step 1.5. Matching equation. The boundary conditions requiring continuity of the logarithmic derivative of the wave function $\phi(y)$ at the phase interface y=0 led to the final transcendental equation:

$$\frac{D_{\nu'_-}(-\sqrt{2}(\alpha+\bar{\lambda}_2))}{D_{\nu_-}(-\sqrt{2}(\alpha+\bar{\lambda}_2))} + \frac{D_{\nu'_+}(\sqrt{2}(\alpha+\bar{\lambda}_2))}{D_{\nu_+}(\sqrt{2}(\alpha+\bar{\lambda}_2))} = 0.$$

2. Probability density distributions $P(\tau_+, I_T)$ and free-energy profiles.

Knowledge of the thermodynamic potential $E(\lambda_1,\lambda_2)$ allows us to take a crucial step regarding the visualization and verification of the theory: calculating the joint probability density function (PDF) of observing a trajectory with specified opening-time and current parameters.

Step 2.1. Relation between the distribution and the potential (Donsker-Varadhan Theory). In accordance with the large deviation principle, the joint PDF at long observation times $T \to \infty$ takes an exponential form: $P(\tau_+, I_T) \sim \exp\left(-T \cdot \mathcal{I}(\bar{\tau}_+, \bar{I})\right)$, where $\bar{\tau}_+ = \tau_+ / T$, $\bar{I} = I_T / T$. The function $\mathcal{I}(\bar{\tau}_+, \bar{I})$ is called the rate function or the dynamic entropy of fluctuations. In a physical context, this is a direct analogue of the non-equilibrium path-space free energy profile.

Step 2.2. Legendre transform. The rate function $\mathcal{I}(\bar{\tau}_+, \bar{I})$ is related to the potential $E(\lambda_1,\lambda_2)$ we calculated via a two-dimensional Legendre transform:

$$\mathcal{I}(\bar{\tau}_+, \bar{I}) = \max_{\lambda_1,\lambda_2}\left[-\lambda_1\bar{\tau}_+ - \lambda_2\bar{I} - E(\lambda_1,\lambda_2)\right].$$

Since the potential $E(\lambda_1,\lambda_2)$ is a quadratic form in the quadratic (Gaussian) approximation, the maximization is performed analytically. By substituting the expansion for the potential and solving the system of linear extremum equations, we find that the rate function is itself a quadratic form determined by the inverse susceptibility matrix $\mathcal{I}(\overline{\tau}_+, \overline{I}) = \frac{1}{2\det\chi}\left[\chi_{22}(\overline{\tau}_+ - \langle\tau_+\rangle)^2 + \chi_{11}(\overline{I} - \langle I\rangle)^2 - 2\chi_{12}(\overline{\tau}_+ - \langle\tau_+\rangle)(\overline{I} - \langle I\rangle)\right]$ (fluctuation matrix), where $\det\chi = \chi_{11}\chi_{22} - \chi_{12}^2$.

Step 2.3. Topography of free-energy profiles and graphical representation (Fig. 3). Plotting the level lines (isolines) of the rate function $\mathcal{I}(\overline{\tau}_+, \overline{I}) = \mathbf{const}$ in the plane of the variables $(\overline{\tau}_+, \overline{I})$ yields a family of nested fluctuation ellipses.

1. Distribution center: The minimum of the rate function $\mathcal{I} = 0$ (the maximum of the probability) is located at the point of mean stationary values: $(\langle\tau_+\rangle_{st}, \langle I\rangle_{st})$. At this point, the trajectory's free energy is minimal.
2. Tilt of the ellipses (Onsager effect): The principal axes of the ellipses are rotated by a certain angle $\theta$ relative to the coordinate axes $(\overline{\tau}_+, \overline{I})$. This tilt angle is directly determined by the cross-coefficient $\chi_{12}$.
   - Since $\chi_{12} > 0$, the ellipses are elongated along the positive diagonal. This reflects a strong physical correlation: the longer the channel remains open (as $\overline{\tau}_+$ increases), the greater the total ionic current $\overline{I}$ it allows to pass.
3. Effect of asymmetry $\mathbf{U}_0$:
   - With an increase in the base channel-blocking level ($U_0 \to \infty$) the center of the ellipse shifts to the left and down (towards small opening times and small currents).
   - At the same time, due to the multiplier $\exp(-\alpha^2)$ in $\chi_{12}$ the fluctuation ellipses align parallel to the current axis. Cross-correlation vanishes. The system may exhibit current fluctuations (due to dispersion within the closed state), but these fluctuations cease to be linked to the gate opening time, since the channel gate virtually never reaches the zero threshold.

Near the distribution center, fluctuations are Gaussian in nature, and the isolines form classic Onsager ellipses. However, as the boundaries $\overline{\tau}_+ \to 0$ and $\overline{\tau}_+ \to 1$ are approached, kinetic asymmetry of the phase space manifests itself. The level lines become distorted, reflecting the fact that the extreme suppression of the channel's open-state duration imposes a strict upper limit on its integral conductance, thereby completely restructuring the topology of the nonequilibrium free-energy potential.

Analyzing the asymptotic behavior of the rate function (the free-energy profile of trajectories) $\mathcal{I}(\overline{\tau}_+, \overline{I})$ at the boundaries of the domain is a crucial part of the physical discussion. At the edges of the grid $\tau_+ \to 0$— where the channel is virtually always closed or constantly open $\tau_+ \to 1$—the Gaussian approximation breaks down completely. Here, strong nonlinear large-deviation effects emerge, driven by strict kinetic constraints.

1. Asymptotic as $\tau_+ \to 0$. (Deep suppression regime). This regime describes extremely rare trajectories in which the channel failed to open almost even once throughout the entire observation period T.

- Behavior of the conjugate force: To force the system to follow such a trajectory, the conjugate force $\lambda_1$ (opening penalty) must tend toward positive infinity $\overline{\lambda}_1 \to +\infty$.
- Asymptotic of the rate function: In this region, the rate function ceases to be quadratic in the variable $\tau_+$. The probability of spending an infinitesimally small amount of time in the open state for diffusion processes is described by "short-time" fluctuation-type asymptotic $\mathcal{I}(\overline{\tau}_+, \overline{I}) \propto \frac{C(\overline{I})}{\overline{\tau}_+^2}$.

Physical meaning and geometry of isolines: The free-energy contour lines on Mathcad plot will begin to crowd together sharply as they approach the vertical axis $\overline{\tau}_+ = 0$. The ellipses deform, transforming into "droplets" elongated along the current axis.

- Physical consequence: Even if the channel is closed ($\tau_+ \to 0$), the integrated current $\overline{I}$ can still fluctuate, as the Brownian particle undergoes oscillations within the closed half-plane ($X_t$

< 0). However, due to the hard wall at zero, current fluctuations become completely symmetric and independent of the gating mechanism.

Asymptotic as $\overline{\tau}_+ \to 1$. (Continuous activation regime). This case is symmetric to the first and describes trajectories where the channel virtually never closed.

- Conjugate force behavior: A strong attractive field is required in the open state: $\overline{\lambda}_1 \to -\infty$.
- Asymptotic behavior of the speed function: $\mathcal{I}(\overline{\tau}_+, \overline{I}) \propto \dfrac{C'(\overline{I})}{(1-\overline{\tau}_+)^2}, \quad \overline{\tau}_+ \to 1$.
- Physical significance: The level lines are pushed against the right boundary of the plot $\overline{\tau}_+ = 1$. In this regime, the system behaves like a pure, unconstrained Ornstein-Uhlenbeck process shifted into the positive half-plane. Current fluctuations are maximal here, as nothing prevents the coordinate $X_t$ from drifting far to the right, generating anomalously high conductance values.

3. Diagonal constraint (kinetic boundary of trajectories). The most profound nonlinear effect you will observe on the Mathcad plot when moving beyond the regime of small λ is the existence of forbidden zones.

Since our integrated current $I_T = \int_0^T X_t dt$ is strictly linked to the residence time (current accumulates only when the particle is in the corresponding half-plane), there are rigorous mathematical inequalities that the trajectory cannot physically violate. For example:

- If $\tau_+ \to 0$, then the process $X_t$ physically cannot stray far into the positive region. Consequently, the average current $\overline{I}$ on such a trajectory is strictly bounded from above: $\overline{I} \le 0$
- If $\tau_+ \to 1$, the process hardly ever enters the negative zone, which implies a lower bound: $\overline{I} \ge 0$.

Addition of an additional independent force V.

Introducing a diffusive drift (a constant external force/drift) into the Langevin equation qualitatively and quantitatively changes the thermodynamic picture. In fact, in our model of the asymmetric Ornstein–Uhlenbeck process, a constant drift is already present in the form of the parameter $U_0$ (12) $dX_t = (-\gamma X_t - U_0)dt + \sqrt{2D}dW_t$).

However, if diffusive drift is understood as the addition of an extra independent force V (for example, a directed electric field pulling ions through the channel), this leads to two fundamental physical effects: the breaking of Onsager symmetry and a dynamic phase transition. Let us examine exactly how this drift alters the topology of the plots in Mathcad.

1. Mathematical modification: Shift of effective forces. If we add a constant drift V to the OU process, the full equation takes the form: $dX_t = (-\gamma X_t - U_0 + V)dt + \sqrt{2D}dW_t$. In dimensionless variables, this is equivalent to a simple renormalization of our asymmetry parameter: $\alpha_{new} = \alpha - \overline{V}$, где $\overline{V} = V / \sqrt{2\gamma D}$.

At first glance, this merely shifts the center of the fluctuation ellipse. However, if the force V acts during only one of the phases (for example, if an electric field pulls ions only when the channel is open—i.e., when $X_t > 0$), the mathematics becomes radically more complex. The potential in the right half-plane acquires an additional slope.

2. Major qualitative changes in the fluctuation pattern. If the drift is asymmetric (depending on the channel state), the following changes will occur in the Mathcad plots:

A. Violation of Onsager reciprocity relations for trajectories. In an equilibrium or purely gradient-driven system (without external drift), the susceptibility matrix is symmetric ($\chi_{12} = \chi_{21}$). The fluctuation ellipses are oriented strictly in accordance with the internal correlation of the process.

- With drift: A constant flow of energy through the system (dissipation) breaks time-reversal symmetry. Cubic and higher-order odd terms appear in the thermodynamic potential $E(\lambda_1, \lambda_2)$.
- On the graph: The fluctuation ellipses do not merely shift; they lose their regular shape (becoming asymmetric "eggs" or "boat shapes"), even near the center of the distribution. Cross-sensitivity $\chi_{12}$ in one direction ceases to equal sensitivity in the other direction.

B. Compression and rotation of the fluctuation ellipse (current stabilization effect). Constant, strong drift $V \gg 0$ begins to "drag" the trajectories across the barrier.

- On the graph: The axis of the fluctuation ellipse begins to rotate and steepen. The current autocorrelation coefficient $\chi_{22}$—which, for a pure OU process, was independent of asymmetry—now begins to decrease (drift narrows the current fluctuation corridor, "pinning" the particle to the directed flow).

B. Possibility of a dynamic phase transition. If the external driving force is sufficiently strong and opposes the oscillator's restoring force, trajectory bistability may arise in the system.

- On the graph: Instead of a single fluctuation center (a single minimum of the velocity function), two independent local minima separated by a saddle point may appear on the contour map. The system splits into two trajectory phases: "fast" (the channel is open, and strong directional drift occurs) and "slow" (the channel is closed, and the system fluctuates due to diffusion).

## Appendix B. Topography of free energy profiles and Onsager ellipses

### 1. Physical Interpretation of the Contour Map

The two-dimensional contour plot of the large deviation rate function M=$\nu_{-}(E,\lambda_2)$ on the $(\lambda_2,E)$ parameter plane (presented in Figure **3**) reveals the deep topological structure of path-space fluctuations in the non-equilibrium system.

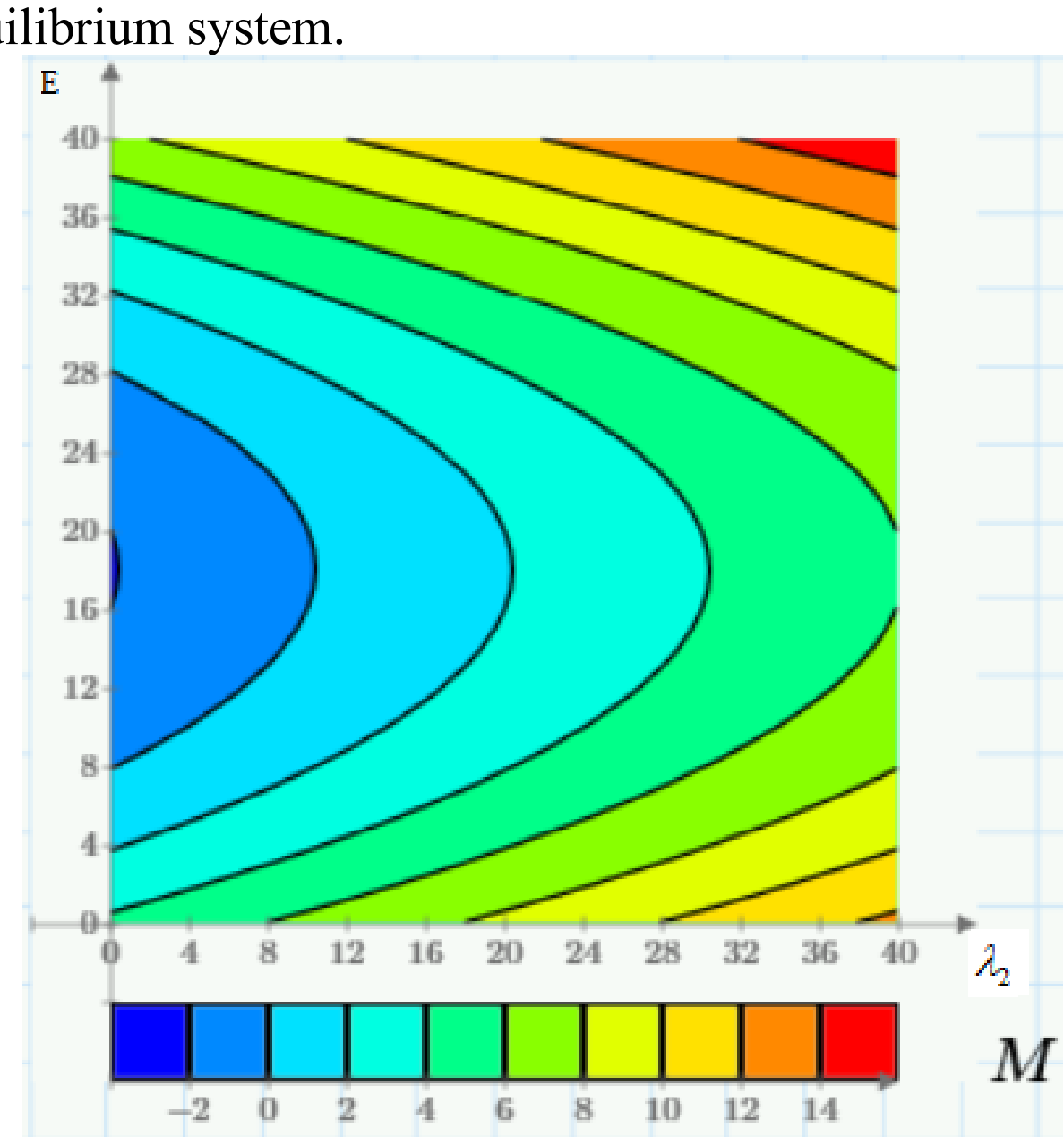

**Figure 3.** Topographic contour map of the path-space free energy profile $\nu_{-}(E,\lambda_2)$. The color gradient corresponds to the magnitude of the rate function matrix $M=\nu_{-}(E,\lambda_2)$. The transition from concentric quadratic ellipses in the linear regime to asymmetric non-parabolic contours at remote boundaries highlights the breakdown of Onsager's near-equilibrium symmetry.

The geometric configuration of the isolines of the rate function matrix $M$ provides a direct path-space visualization of thermodynamic fluctuations. In the immediate vicinity of the distribution center (the minimum energy state where $M\rightarrow-2$), the fluctuations exhibit a strictly Gaussian character. In this near-equilibrium linear regime, the isolines manifest as classic Onsager ellipses, reflecting the quadratic nature of the generalized entropy production and the standard fluctuation-dissipation relationships. The concentric alignment of these ellipses indicates that small path-space deviations obey reciprocal symmetry and linear relaxation kinetics.

However, as the system moves farther from the stationary center toward the remote boundaries of the configuration space, a profound topological distortion of the elliptic contours becomes apparent. The contours systematically lose their idealized quadratic symmetry, tilting and elongating along the energy axis $E$. This geometric deformation explicitly manifests the kinetic asymmetry of the phase space induced by the joint action of the strong directional drift $\lambda_2=-2.5$ and the discontinuous boundary step potential $\lambda_1$.

Far from the equilibrium core, the non-linear coupling between the integral occupation time and extreme boundary events shifts the fluctuation weights. The asymmetric stretching of the outer ellipses proves that high-energy, rare fluctuations do not follow the time-reversible path-space scenarios predicted by linear Onsager kinetics. Instead, they are governed by non-parabolic rate functions, signaling a dynamic phase separation where individual macroscopic trajectories become strictly irreversible.